\documentclass[aps,preprintnumbers,superscriptaddress,showpacs]{revtex4}
\usepackage{epsfig}
\usepackage{psfrag}
\usepackage{amsfonts}
\usepackage{graphicx}
\usepackage{dcolumn}
\usepackage{bm}
\usepackage{amsmath}

\begin{document}

\title{Flavor-resolved screening length of a moving quarkonium in a
lattice-calibrated holographic QCD plasma}

\author{Huwei Zhu}
\email{zhuhuwei@cug.edu.cn} \affiliation{School of Mathematics and
Physics, China University of Geosciences (Wuhan), Wuhan 430074,
China}

\author{Zi-qiang Zhang}
\email{zhangzq@cug.edu.cn} \affiliation{School of Mathematics and
Physics, China University of Geosciences (Wuhan), Wuhan 430074,
China}

\begin{abstract}
We compute the screening length \(L_s\) of a moving heavy
quark-antiquark pair in a lattice-calibrated
Einstein-Maxwell-dilaton background for pure-glue, two-flavor, and
\(2+1\)-flavor matter. Parallel and perpendicular orientations are
considered at finite temperature, baryon chemical potential, and
rapidity. \(L_s\) decreases monotonically with \(T\) and \(\eta\)
in every flavor sector, and for the flavored systems also with
\(\mu\). The perpendicular screening length is systematically
smaller than the parallel one. At zero density and \(T=0.3\
\mathrm{GeV}\), the ordering
\(L_s^{(N_f=0)}>L_s^{(N_f=2)}>L_s^{(N_f=2+1)}\) holds in this
calibration, and a lattice check of the static flavor dependence
is proposed. At finite density the two-flavor and \(2+1\)-flavor
curves cross along a line in the \(T\)-\(\mu\) plane; below the
line \(L_s^{(N_f=2)}>L_s^{(N_f=2+1)}\), while above it the
ordering reverses. The crossing follows from the competition
between the two-flavor crossover temperature and the density
coupling, and its location depends on the fitted parameters. These
results provide a flavor-resolved reference for moving quarkonium
screening in the baryon-rich regime probed by beam energy scans.
\end{abstract}

 \maketitle

\section{Introduction}

Ultrarelativistic heavy-ion collisions at the Relativistic Heavy
Ion Collider (RHIC) and the Large Hadron Collider (LHC) have
established the formation of a strongly coupled quark-gluon plasma
(QGP)~\cite{Shuryak:2004,Kovtun:2005}. A key experimental
signature of QGP formation is the suppression of quarkonium states
such as $J/\psi$ and $\Upsilon$, proposed by Matsui and Satz
\cite{Matsui:1986} as a direct probe of color screening in the
deconfined medium. When the medium temperature is sufficiently
high, the inter-quark potential is screened beyond a critical
separation $L_s$, and the quarkonium bound state dissociates. The
dependence of $L_s$ on temperature, baryon density, and quark
flavor content therefore encodes the in-medium interaction
strength and the QCD phase structure.

First-principles calculations of moving quarkonium in a strongly
coupled QGP remain challenging. Lattice QCD is hampered by the
sign problem at finite baryon density and by technical
difficulties in real-time dynamics~\cite{Kaczmarek:2005}, while
perturbative QCD breaks down near the deconfinement transition.
The gauge/gravity duality provides a nonperturbative alternative
by mapping four-dimensional strongly coupled gauge theories to
weakly coupled five-dimensional gravitational
systems~\cite{Maldacena:1997,Gubser:1998,Witten:1998}.

Within the holographic framework, $L_s$ is a useful diagnostic of
QGP properties and offers a natural explanation for $J/\psi$
suppression. Liu, Rajagopal, and Wiedemann first computed $L_s$
for a heavy quark-antiquark ($Q\bar{Q}$) pair moving in
$\mathcal{N}=4$ super Yang-Mills (SYM)
plasma~\cite{Liu:2007,Liu:2007b}. They considered a fundamental
Wilson loop describing a color-singlet dipole moving at constant
speed and identified $L_s$ with the maximum separation for which a
connected string solution exists. For $L>L_s$, the quarks lose
their binding energy and are screened by the plasma. The same
Wilson loop also yields the binding energy from its thermal
expectation value~\cite{Maldacena:1998,Rey:1998,Brandhuber:1998}.
Subsequent studies extended the holographic screening length to
nonconformal and anisotropic geometries, quark-monopole
configurations, and bottom-up AdS/QCD
models~\cite{Natsuume:2007,Caceres:2006,Chernicoff:2006,Athanasiou:2008,Chernicoff:2008,Liu:2008,Krishnan:2008,Ejaz:2008,Sin:2009,Paredes:2009,Sadeghi:2010,Chernicoff:2013,Cai:2013,Atmaja:2015,Xu:2015,Chakrabortty:2016,Sadofyev:2016,Zhu:2022,Zhu:2024,Zhang:2024,Wu:2024,Wu:2025,Pokhrel:2025a,Pokhrel:2025b,Pokhrel:2026}.
Many of these calculations either use conformal or non-QCD
backgrounds or treat the pair at rest; comparatively few combine a
lattice-calibrated nonconformal background with finite velocity,
finite density, and multiple flavors.

Experimentally, beam-energy scans at RHIC show nontrivial
variations of quarkonium yields with collision energy, indicating
sensitivity to the QCD critical endpoint and finite baryon
density~\cite{Andronic:2016,Andronic:2024,Veres:2019}.
Measurements of strange-hadron and charmonium suppression in
different collision systems also suggest that dynamical quark
flavors modify the in-medium color screening strength. These
features are not captured by minimal conformal holographic models
and motivate more realistic QCD-like gravitational backgrounds.

To go beyond the conformal SYM paradigm, improved holographic QCD
models have been developed that incorporate nonconformal effects,
running coupling, and the confinement-deconfinement
transition~\cite{DeWolfe:2011,Gursoy:2008a,Gursoy:2008b,Arefeva:2021,Rougemont:2020,Critelli:2017,Rougemont:2024}.
Chen \textit{et al.} constructed an Einstein-Maxwell-dilaton (EMD)
framework whose thermodynamics is calibrated to lattice QCD data
through machine-learning
optimization~\cite{Chen:2024,Chen:2025,Zeng:2026}. The framework
encodes the effects of dynamical light and strange quarks in
flavor-dependent bulk parameters and yields distinct zero-density
crossover temperatures and critical endpoints for different flavor
contents. It has been applied to parton energy loss and jet
quenching~\cite{Zhang:2025,Zhu:2026,Ficnar:2013,Chesler:2014,Rajagopal:2016,Casalderrey:2014,Chenb:2025,Chenb:2026,Guo:2024,Zhulq:2026,Zhou:2026,Zhuzr:2026,zhulq:2025,zhulq:2026b,lin:2025,liu:2026,changwb:2026}.

Color screening of a heavy quark-antiquark pair is among the few
in-medium quantities accessible both on the lattice and
holographically. In $\mathcal{N}=4$ SYM theory, the velocity
dependence is known analytically in the ultrarelativistic
limit~\cite{Liu:2007,Caceres:2006}, the angular dependence has
been established~\cite{Natsuume:2007}, and the framework has been
extended to baryons~\cite{Athanasiou:2008}, accelerating
pairs~\cite{Chernicoff:2008}, and various deformed backgrounds,
including curvature-squared corrections~\cite{Zhang:2024},
soft-wall geometries~\cite{Zhu:2024}, noncommutative
spaces~\cite{Wu:2024}, and magnetized plasmas~\cite{Wu:2025}.

Many of these calculations either use conformal or non-QCD
backgrounds or treat the pair at rest. The EMD framework offers a
nonconformal alternative~\cite{Cai:2013}, but its screening
lengths have in many cases lacked lattice QCD calibration. The
static case in a lattice-calibrated EMD background was recently
studied for the 2+1-flavor system~\cite{Zhuzr:2026}. The situation
relevant for collisions, however, involves a quarkonium state
moving with finite velocity relative to the fluid: the wind
modifies the screening length, and the angular dependence
distinguishes pairs aligned with the wind from those transverse to
it. At beam-energy-scan energies, the medium is also
flavor-resolved and at finite baryon chemical potential. Because
the three flavor systems have different crossover temperatures and
equations of state, their screening lengths need not remain
universally ordered across the $T$-$\mu$ plane.

This paper addresses this setting. Using the machine-learning
calibrated EMD metric of Ref.~\cite{Chen:2024}, whose parameters
are fixed by the lattice equation of state and baryon
susceptibility, we compute $L_s$ for a moving pair in three flavor
contents, at finite $\mu$ for the flavored systems, and for
parallel and perpendicular orientations. Because the metric is
thermodynamically calibrated, $L_s$ is a prediction with no
additional parameters beyond those fixed by thermodynamics, so its
flavor and density dependence provides a genuine test of the
model. We find that the flavor ordering is not universal: it
reverses along a line in the $(T,\mu)$ plane, a structure absent
in conformal treatments. The same background has been used for jet
quenching~\cite{Zhang:2025,Zhu:2026}, which allows the two probes
to be compared at fixed absolute temperature $T$ or at fixed
reduced temperature $T/T_c$; these two schemes order the media
differently.

The remainder of this paper is organized as follows.
Sec.~\ref{sec:background} reviews the five-dimensional EMD
framework~\cite{Chen:2024,Chen:2025}. In Sec.~\ref{sec:screening},
we investigate the screening length of a heavy $Q\bar{Q}$ pair
moving in this background and consider the axis of the $Q\bar{Q}$
pair along different directions. Sec.~\ref{sec:results} presents
the numerical results for both orientations, discussing the
rapidity dependence, the flavor ordering at zero density, the
effects of finite temperature and baryon chemical potential, and
the resulting flavor-inversion line. Our conclusion and outlook
are presented in Sec.~\ref{sec:conclusion}.

\section{Background geometry}
\label{sec:background}

Color screening and quarkonium dissociation are tied to the
thermodynamic and geometric properties of the deconfined medium. A
reliable holographic description therefore requires a
gravitational dual that reproduces the equation of state, the
confinement-deconfinement crossover, and the finite-density
response of real QCD. The Einstein-Maxwell-dilaton (EMD) framework
is well suited to this task: its dilaton and gauge fields provide
enough parametric freedom to encode nonconformal and
finite-density effects while remaining tractable for explicit
string calculations.

We adopt the same five-dimensional EMD action employed in the
lattice-calibrated holographic QCD construction of
Refs.~\cite{Chen:2024,Chen:2025}, whose bulk parameters were
determined via machine-learning optimization against lattice QCD
thermodynamics for pure-glue, two-flavor, and 2+1-flavor media. In
the Einstein frame, the gravitational action reads
\begin{equation}
S = \int \frac{d^5 x}{16\pi G_5} \sqrt{-\det(g_{\mu\nu})} \left[R
- \frac{f(\phi)}{4} F_{\mu\nu}F^{\mu\nu} - \frac{1}{2}
\partial_\mu \phi \partial^\mu \phi - V(\phi)\right],
\end{equation}
where $G_5$ denotes the five-dimensional Newton constant, $R$ the
Ricci scalar, $F_{\mu\nu}=\partial_\mu A_\nu-\partial_\nu A_\mu$
the Maxwell field strength dual to the boundary baryon number
current, $\phi$ the dilaton field, $f(\phi)$ the dilaton-dependent
gauge kinetic function, and $V(\phi)$ the dilaton potential.
Varying this action yields a coupled system of Einstein, dilaton,
and Maxwell equations, whose closed-form solutions follow once a
suitable warped black brane metric is specified.

We use the standard static, homogeneous, and isotropic black-brane
ansatz for finite-density dilaton holography,
\begin{equation}
ds^2 = \frac{R^2 e^{2A(z)}}{z^2} \left[ -g(z) dt^2 + d\vec{x}^2 +
\frac{dz^2}{g(z)} \right], \label{ansatz}
\end{equation}
with $R$ the AdS curvature radius and $z$ the fifth-dimensional
radial coordinate. The boundary is at $z\to 0$, while the
black-brane horizon is at finite $z=z_t$, where $g(z_t)=0$. The
warp factor $A(z)$ incorporates the nonconformal corrections
inherited from boundary QCD, and the blackening function $g(z)$,
subject to the horizon regularity condition $g(z_t)=0$, governs
the bulk causal structure that determines the maximum radial
penetration of holographic Wilson loops and fundamental strings.
The temporal bulk gauge field $A_t(z)$ encodes the boundary baryon
chemical potential $\mu$ through the conditions $A_t(0)=\mu$ and
$A_t(z_t)=0$.

Adopting the machine-learning fitting procedure detailed in
Ref.~\cite{Chen:2024}, we obtain the following closed-form
analytic expressions for $g(z)$, $\phi'(z)$, $A_t(z)$, and $V(z)$:
\begin{align}
g(z) &= 1 - \frac{1}{\displaystyle\int_{0}^{z_t}dx \, x^3 e^{-3A(x)}} \left[ \int_{0}^{z} dx \, x^3 e^{-3A(x)} + \frac{2c\mu^2 e^k}{(1 - e^{-c z_t^2})^2} \det \mathcal{G} \right], \\
\phi'(z) &= \sqrt{6\left(A'^2 - A'' - 2A'/z\right)},  \label{phi z prime} \\
A_t(z) &= \mu \frac{e^{-c z^2} - e^{-c z_t^2}}{1 - e^{-c z_t^2}}, \\
V(z) &= -\frac{3 z^2 g e^{-2A}}{R^2} \left[ A'' + A'\left(3A' -
\frac{6}{z} + \frac{3g'}{2g}\right) -
\frac{1}{z}\left(-\frac{4}{z} + \frac{3g'}{2g}\right) +
\frac{g''}{6g} \right],
\end{align}
where the determinant $\det\mathcal{G}$ is defined via the
$2\times2$ integral matrix
\begin{equation}
\det \mathcal{G} =
\begin{vmatrix}
\displaystyle\int_{0}^{z_t} dy\, y^3 e^{-3A(y)} & \displaystyle\int_{0}^{z_t} dy\, y^3 e^{-3A(y)-c y^2} \\
\displaystyle\int_{z_t}^{z} dy\, y^3 e^{-3A(y)} &
\displaystyle\int_{z_t}^{z} dy\, y^3 e^{-3A(y)-c y^2}
\end{vmatrix}.
\end{equation}

The Hawking temperature \cite{Hawking:1983,Natsuume:2015} of this
black-hole solution is
\begin{align}
T &= \frac{z_t^3 e^{-3A(z_t)}}{4\pi \displaystyle\int_{0}^{z_t}
dy\, y^3 e^{-3A(y)}} \left[ 1 + \frac{2c\mu^2 e^k \left( e^{-c
z_t^2} \displaystyle\int_{0}^{z_t} dy\, y^3 e^{-3A(y)} -
\displaystyle\int_{0}^{z_t} dy\, y^3 e^{-3A(y)-c y^2} \right)}{(1
- e^{-c z_t^2})^2} \right].
\end{align}

In this calibrated EMD framework, the warp factor takes the form
$A(z) = d \ln(a z^2 + 1) + d \ln(b z^4 + 1)$, and the
dilaton-dependent gauge kinetic function is $f(z) = e^{c z^2 -
A(z) + k}$. For $N_f = 0$, $2$, and $2+1$, the six free
coefficients $\{a, b, c, d, k, G_5\}$ are optimized by gradient
descent against lattice QCD equations of state and baryon
susceptibility~\cite{Chen:2025}. The fitted parameters and the
zero-density crossover temperatures $T_c$ are listed in
Table~\ref{tab:emd_params}. For pure glue ($N_f=0$), $c=0$ removes
the baryon chemical potential dependence from the metric
functions, as expected for Yang-Mills theory without quark or
baryon charge.

\begin{table}[htbp]
  \centering
  \begin{tabular}{|c|c|c|c|c|c|c|c|}
    \hline
     & $a$ & $b$ & $c$ & $d$ & $k$ & $G_5$ & $T_c$ \\
    \hline
    $N_f = 0$ & $0$ & $0.072$ & $0$ & $-0.584$ & $0$ & $1.326$ & $0.265$ \\
    \hline
    $N_f = 2$ & $0.067$ & $0.023$ & $-0.377$ & $-0.382$ & $0$ & $0.885$ & $0.189$ \\
    \hline
    $N_f = 2 + 1$ & $0.204$ & $0.013$ & $-0.264$ & $-0.173$ & $-0.824$ & $0.400$ & $0.128$ \\
    \hline
  \end{tabular}
  \caption{Bulk parameters from the machine-learning fits of
  Ref.~\cite{Chen:2024}. Units: $G_5$ in $\text{GeV}^3$, $a$ and $c$
  in $\text{GeV}^2$, and $b$ in $\text{GeV}^4$. $T_c$ denotes the
  zero-density crossover temperature for each flavor plasma,
  extracted from lattice-calibrated thermodynamics.}
  \label{tab:emd_params}
\end{table}

The three flavor configurations have distinct significance for
quarkonium phenomenology. $N_f=0$ corresponds to pure Yang-Mills
theory and provides a baseline for isolating dynamical-quark
effects on color screening. $N_f=2$ includes the light $u$ and $d$
quarks and approximates the light-quark sector relevant to
charmonium suppression. $N_f=2+1$ further includes the strange
quark and gives the most realistic description of the QGP produced
in heavy-ion collisions, where strangeness affects the medium
properties. As validated in the original
construction~\cite{Chen:2024} and in jet-quenching
studies~\cite{Zhang:2025,Zhu:2026}, this six-parameter setup
reproduces lattice QCD thermodynamics, including pressure, energy
density, and baryon susceptibility, for all three sectors. It also
yields distinct critical endpoints and chemical-potential
trajectories for $N_f=2$ and $N_f=2+1$ QCD~\cite{Chen:2025},
reflecting how dynamical flavors reshape the phase diagram and the
screening environment.

With these lattice-calibrated bulk geometries, we proceed in the
next section to construct the fundamental-string configurations
that encode the screening length of a moving $Q\bar{Q}$ pair.

\section{Screening length in the EMD background}
\label{sec:screening}

We work in the string frame, where the fundamental string couples
minimally to the metric. The string-frame warp factor is related
to the Einstein-frame one by \(A_s(z)=A(z)+\phi(z)/\sqrt{6}\),
with \(\phi(z)\) obtained by integrating Eq.~\eqref{phi z prime}.
We set \(R=1\) and define
\begin{equation}
E(z)\equiv \frac{e^{2A_s(z)}}{z^2}.
\end{equation}
The Wilson loop also determines the in-medium quark-antiquark
potential and the dissociation criterion
\cite{Maldacena:1998,Rey:1998,Brandhuber:1998}. A pair moving with
rapidity \(\eta\) along \(x_3\) is described by boosting the
metric~\cite{Liu:2007},
\begin{equation}
dt = dt'\cosh\eta - dx_3'\sinh\eta, \qquad dx_3 = -dt'\sinh\eta +
dx_3'\cosh\eta,
\end{equation}
where \(\sinh\eta=v/\sqrt{1-v^2}\). After dropping the primes, the
boosted string-frame metric reads
\begin{align}
ds_s^2 &= \frac{e^{2A_s(z)}}{z^2} \biggl\{
\Bigl[-g(z)\cosh^{2}\eta+\sinh^{2}\eta\Bigr]\,dt^{2}
+\Bigl[-g(z)\sinh^{2}\eta+\cosh^{2}\eta\Bigr]\,dx_{3}^{2} \notag \\
&\quad+2\sinh\eta\cosh\eta\,[g(z)-1]\,dt\,dx_{3}
+dx_{1}^{2}+dx_{2}^{2}+\frac{dz^{2}}{g(z)} \biggr\}.
\end{align}
The off-diagonal \(dt\,dx_3\) term is the origin of the
velocity-induced anisotropy between the parallel and perpendicular
screening lengths. The classical string dynamics follows from the
Nambu-Goto action
\begin{equation}
S_{NG} = -\frac{1}{2\pi\alpha'}\int d\tau d\sigma \sqrt{-\det
g_{\alpha\beta}}, \qquad g_{\alpha\beta}=G_{\mu\nu}\frac{\partial
X^\mu}{\partial\sigma^\alpha}\frac{\partial
X^\nu}{\partial\sigma^\beta},
\end{equation}
with \(\alpha'\) the Regge slope. For the configurations
considered below, the Lagrangian density does not depend
explicitly on \(\sigma\), so there is a conserved quantity
\begin{equation}
\mathcal{L}-\frac{\partial\mathcal{L}}{\partial\dot{z}}\dot{z}=M,
\qquad \dot{z}\equiv\frac{dz}{d\sigma},
\end{equation}
which parametrizes the family of U-shaped string solutions and
fixes the endpoint separation on the boundary. The constant \(M\)
labels the turning point through the condition \(\dot{z}=0\); the
maximum of the endpoint separation over this family defines the
screening length.

\subsection{Orientation and wind anisotropy}
\label{subsec:orientation}

The boost above selects \(x_3\) as the wind direction. Before the
boost the EMD metric is spatially isotropic in \(x_1,x_2,x_3\);
after the boost it is invariant under rotations in the transverse
\(x_1\)-\(x_2\) plane. Thus the only anisotropy is the wind, and
the orientation of the \(Q\bar{Q}\) axis relative to the wind is
specified by a single angle \(\theta\) in the \(x_3\)-\(x_1\)
plane. The parallel orientation corresponds to \(\theta=0\), and
the perpendicular orientation to \(\theta=\pi/2\). Because of the
residual \(SO(2)\) symmetry, all transverse directions are
equivalent, so there is no additional independent perpendicular
case. This differs from the noncommutative plasma of
Ref.~\cite{Wu:2024}, where the noncommutative plane breaks the
spatial rotation symmetry and several orientations relative to the
noncommutative directions must be distinguished. In the present
EMD background the wind is the only source of anisotropy, and the
two limiting orientations are the principal axes of the
wind-induced effective geometry.

For a general angle \(0<\theta<\pi/2\), the worldsheet ansatz must
mix \(x_1\) and \(x_3\), and \(L_s(\theta)\) is not obtained by
interpolating the two limits. Here we compute the two extreme
orientations, which provide the reference endpoints of the
wind-induced splitting and are sufficient for the statements made
below: the ordering \(L_{s\perp}<L_{s\parallel}\), its persistence
across flavor content, and, for the flavored systems where a
chemical potential dependence is present, the monotonic dependence
on \(T\), \(\mu\), and \(\eta\). A complete angular scan,
including possible nonmonotonicity of \(L_s(\theta)\), is left for
future work. This treatment parallels the hot-wind calculations of
Refs.~\cite{Liu:2007,Natsuume:2007}. In the magnetized calculation
of Ref.~\cite{Wu:2025} the magnetic field selects a preferred
direction and the orientation dependence is controlled by the
field; in the present EMD background the wind is the only source
of anisotropy, so the parallel and perpendicular cases are the two
independent limits.

\subsection{Parallel orientation}

For the parallel orientation, the \(Q\bar{Q}\) axis lies along the
direction of motion. We parametrize the worldsheet as
\begin{equation}
\tau=t,\qquad \sigma=x_3,\qquad x_1=x_2=0,\qquad z=z(\sigma),
\end{equation}
with boundary conditions
\begin{equation}
z(0)=z_c,\qquad z(\pm L/2)=0,\qquad \dot{z}(0)=0,
\end{equation}
where \(z_c\) is the turning point and \(L\) the coordinate
endpoint separation. The induced Lagrangian density becomes
\begin{equation}
\mathcal{L}=E(z)\sqrt{g(z)+b(z)\dot{z}^{2}}, \qquad b(z)\equiv
\cosh^{2}\eta-\frac{\sinh^{2}\eta}{g(z)}.
\label{eq:Lparallel-Lagrangian}
\end{equation}
Combining the conserved quantity with
Eq.~\eqref{eq:Lparallel-Lagrangian} gives
\begin{equation}
\dot{z}^{2}=\frac{g(z)\left[E^{2}(z)g(z)-M^{2}\right]}{b(z)M^{2}}.
\end{equation}
At the turning point \(\dot{z}=0\), which gives
\begin{equation}
E^{2}(z_{c})g(z_{c})=M^{2}.
\end{equation}
Integrating from the boundary to the turning point, the parallel
screening distance is
\begin{equation}
L_{\parallel}=2M\int_{0}^{z_{c}}dz\,\sqrt{
\frac{b(z)}{a_1^2(z)-g(z) M^{2}}}, \qquad a_1(z)=E(z)g(z).
\label{eq:Lparallel}
\end{equation}
For fixed medium parameters \((T,\mu,\eta)\) and flavor content,
Eqs.~\eqref{eq:Lparallel} and the turning-point condition
determine \(L_\parallel\) as a function of \(M\). The maximum of
\(L_\parallel(M)\) over all physically allowed \(M\) defines the
parallel screening length \(L_{s\parallel}\), the critical
separation beyond which no connected string solution exists.

\subsection{Perpendicular orientation}

For the perpendicular orientation, the \(Q\bar{Q}\) axis is
transverse to the direction of motion. The worldsheet
parametrization is
\begin{equation}
\tau = t,\qquad \sigma = x_1,\qquad x_2 = x_3 = 0,\qquad z =
z(\sigma).
\end{equation}
It is useful to define
\begin{equation}
h(z)\equiv g(z)\cosh^{2}\eta-\sinh^{2}\eta,
\end{equation}
so that \(b(z)=h(z)/g(z)\). The induced Lagrangian density is
\begin{equation}
\mathcal{L}=E(z)\sqrt{h(z)+b(z)\dot{z}^{2}}.
\end{equation}
The turning-point condition now gives
\begin{equation}
E^{2}(z_{c})h(z_{c})=M^{2}.
\end{equation}
The perpendicular screening distance is
\begin{equation}
L_{\perp}=2M\int_{0}^{z_{c}}dz\,\sqrt{\frac{b(z)}{a_2^2(z)-h(z)M^{2}}},
\qquad a_2(z)=E(z)h(z). \label{eq:Lperp}
\end{equation}
Note that \(b(z)\) is common to both orientations, while the
effective radial function differs: it is proportional to \(g(z)\)
in the parallel case and to \(h(z)\) in the perpendicular case.
For the perpendicular orientation, \(h(z)\) vanishes at a finite
radial position \(z_0<z_t\) defined by \(g(z_0)=\tanh^2\eta\). The
function \(b(z)\) also changes sign at \(z_0\), so neither string
penetrates beyond this point. The difference between the two
orientations comes from the turning-point condition and from the
effective radial function in the integrand.

Both orientations share the function \(b(z)\), which changes sign
at \(z_0\). In the perpendicular case the turning-point condition
involves \(h(z)\), and \(h(z)<g(z)\) for \(\eta>0\); the effective
radial function is therefore reduced relative to the parallel
case. This reduces the maximum endpoint separation for the
transverse pair. The ordering \(L_{s\perp}<L_{s\parallel}\) is the
same wind-induced effect found in
Refs.~\cite{Liu:2007,Natsuume:2007}; it follows from the boosted
causal structure rather than from the detailed warp factor. The
magnetic calculation of Ref.~\cite{Wu:2025} also finds a strong
orientation dependence, but there the anisotropy is due to the
magnetic field, so the two orderings should not be equated.

\section{Numerical results}
\label{sec:results}

\subsection{Numerical procedure}

The screening length is obtained by evaluating
Eqs.~\eqref{eq:Lparallel} and~\eqref{eq:Lperp} on a grid of
turning points \(z_c\) for each parameter set, locating the
maximum of \(L(M)\), and refining the maximum by local
interpolation. The integrals are computed with adaptive
quadrature. The integrands have integrable square-root
singularities at the turning point, which are handled by the
adaptive step-size routine. A detailed uncertainty analysis is
left for future work. In the static limit, the expressions reduce
to the standard Wilson-loop result, providing a consistency check.
We present results for \(\eta=0.1\) and \(0.3\). At \(\eta=0.1\)
the anisotropy is at the percent level; at \(\eta=0.3\) it is
visibly larger.

The results fall into two classes. For the flavored systems, the
monotonic \(T\), \(\mu\), and \(\eta\) dependences and the
ordering \(L_{s\perp}<L_{s\parallel}\) follow from the horizon
structure and the wind-induced effective geometry. The
zero-density flavor ordering and the finite-density flavor
inversion line depend on the calibrated parameters. We discuss
each group in turn.

\subsection{Rapidity dependence and flavor ordering at zero density}
\label{subsec:zero_density}

\begin{figure}[htbp]
    \centering
    \includegraphics[width=5.9cm]{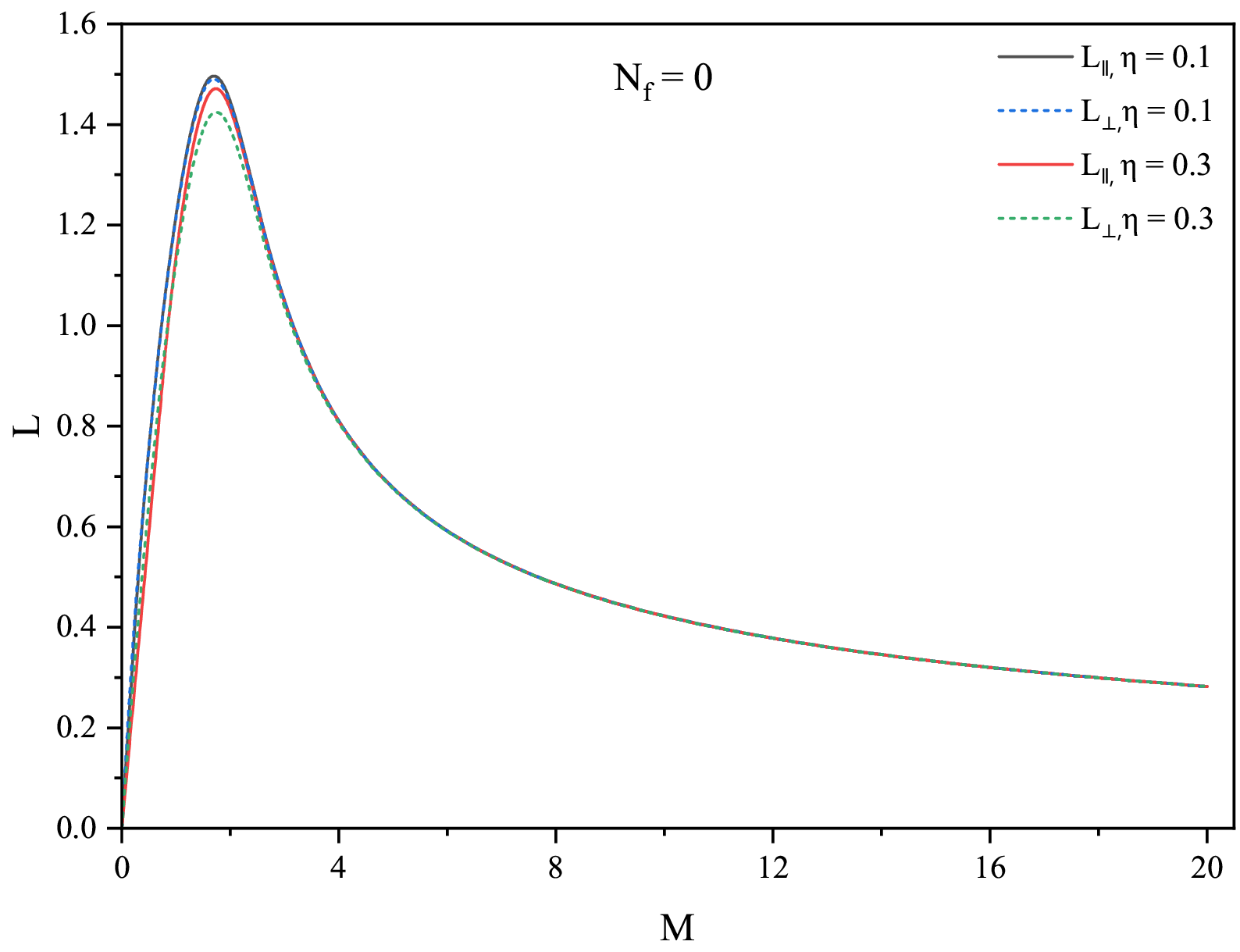}
    \includegraphics[width=5.9cm]{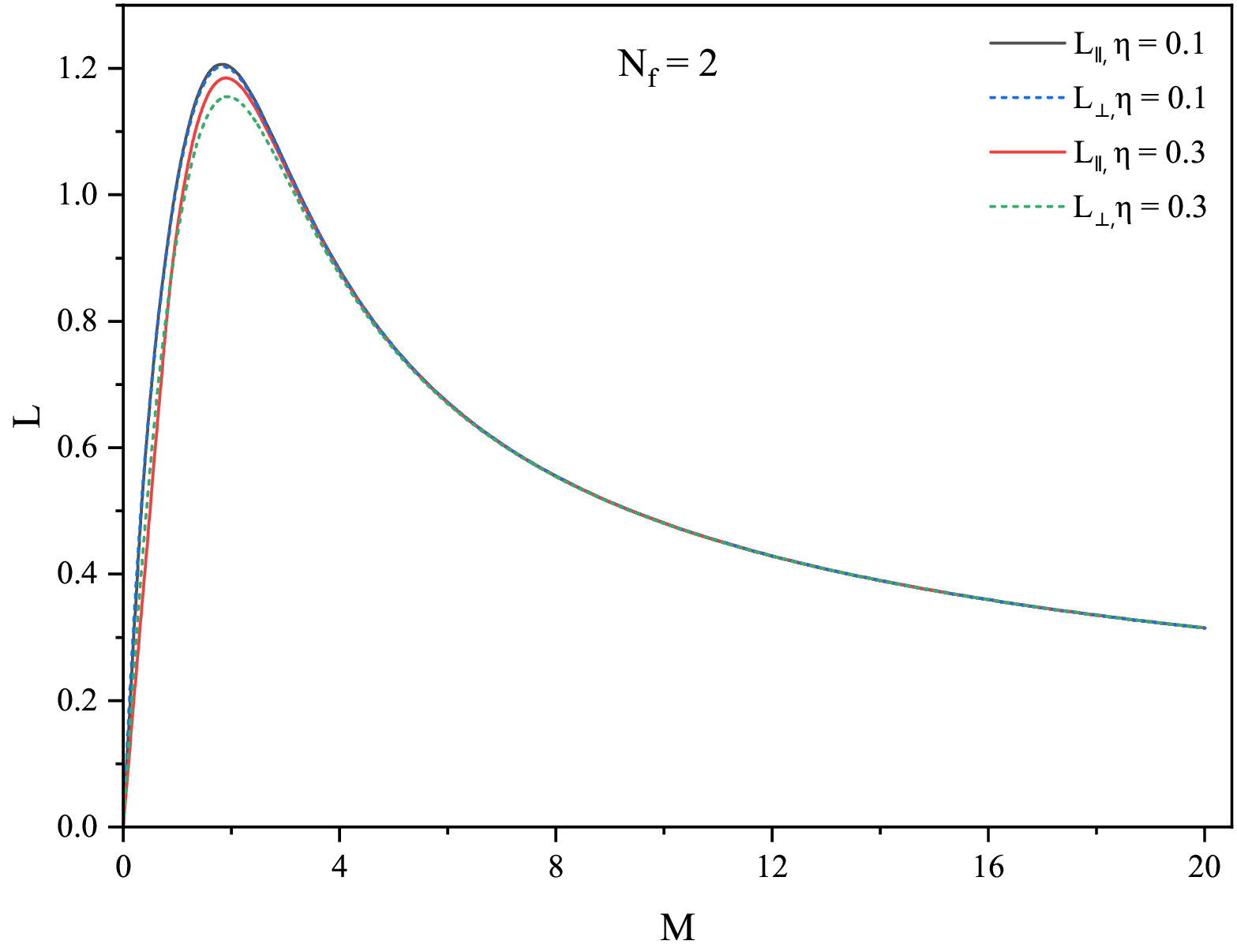}
    \includegraphics[width=5.9cm]{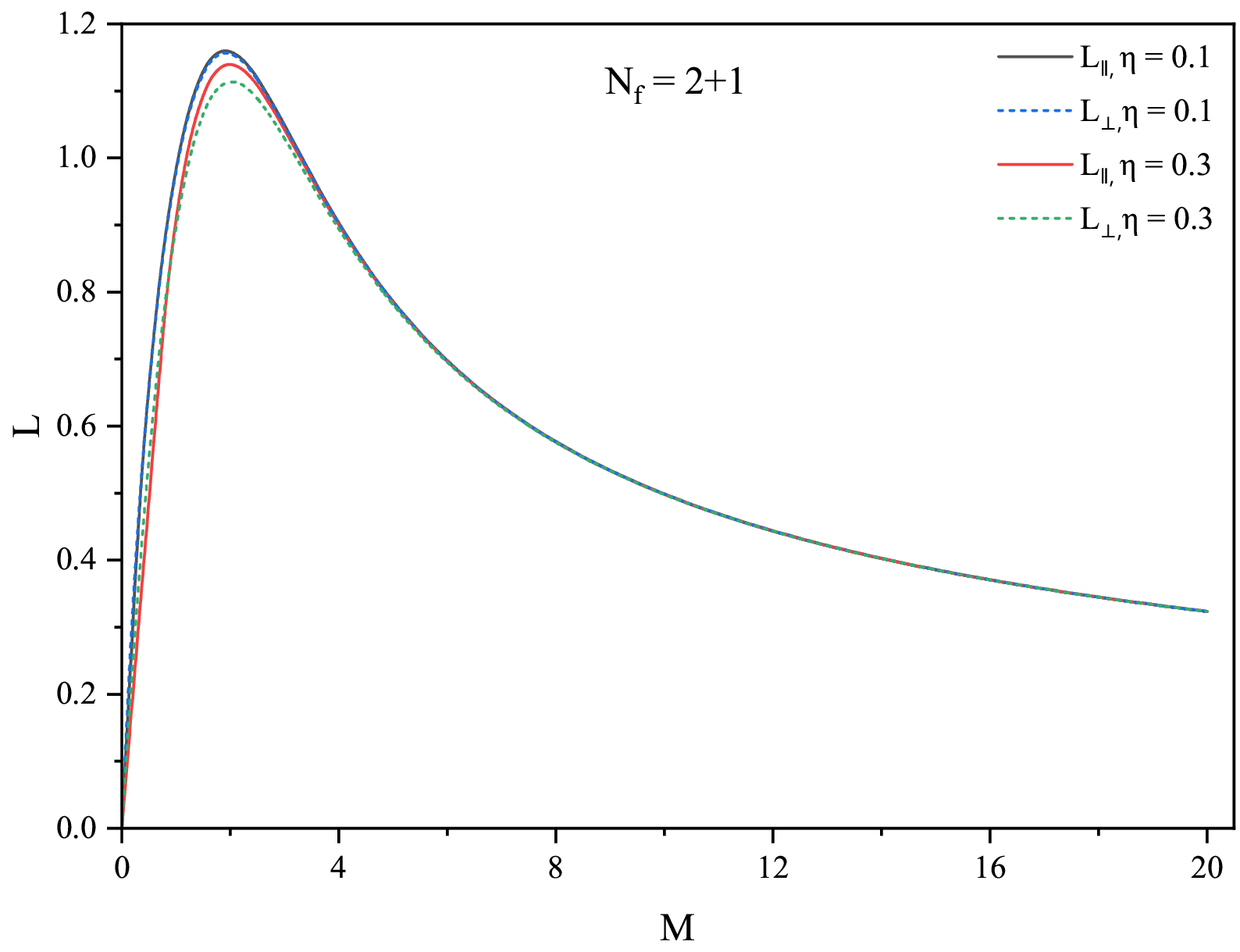}
    \caption{\(L\) versus \(M\) at \(\mu=0\) and
    \(T=0.3\ \text{GeV}\) for \(\eta=0.1\) and \(0.3\). Solid and
    dashed lines denote the parallel and perpendicular orientations.
    Panels: \(N_f=0\) (left), \(N_f=2\) (middle), \(N_f=2+1\)
    (right). The peak of each curve defines \(L_s\); the
    perpendicular peak is systematically lower at fixed \(\eta\), and
    both peaks decrease with increasing \(\eta\).}
    \label{fig1}
\end{figure}

Fig.~\ref{fig1} shows \(L(M)\) for the three flavor systems at
\(\mu=0\), \(T=0.3\ \text{GeV}\), and rapidities \(\eta=0.1\) and
\(0.3\), with solid (dashed) lines for the parallel
(perpendicular) orientation.

At fixed flavor content, increasing \(\eta\) lowers the peak of
\(L\), qualitatively consistent with the conformal scaling found
in Ref.~\cite{Liu:2007}. The perpendicular screening length is
smaller than the parallel one, as expected from the turning-point
condition discussed in Sec.~\ref{subsec:orientation}: the
effective function \(h(z)\) is smaller than \(g(z)\), which
reduces the maximum endpoint separation. Both orientations share
\(b(z)\), which changes sign at \(z_0\), so neither string
penetrates beyond that point; the ordering comes from the
different effective radial functions, not from a different horizon
reach. This is qualitatively similar to the magnetic case of
Ref.~\cite{Wu:2025}, where an anisotropic medium reduces \(L_s\)
most strongly along the directions most affected by the
anisotropy, although the origin of the anisotropy differs.

At \(\mu=0\) and \(T=0.3\ \text{GeV}\), the ordering is
\(L_s^{(N_f=0)}>L_s^{(N_f=2)}>L_s^{(N_f=2+1)}\) for both
orientations: in this calibration additional dynamical flavors
shorten \(L_s\). This ordering at zero density is a prediction of
the calibrated background. It is not a direct consequence of the
crossover-temperature hierarchy alone, since at fixed \(T=0.3\
\text{GeV}\) all three systems lie above their respective
crossovers; it reflects the combined effect of the different warp
factors and the different values of \(G_5\). A lattice
determination of the static flavor dependence, in the spirit of
Ref.~\cite{Kaczmarek:2005}, would test this prediction.

\subsection{Finite baryon chemical potential and temperature}
\label{subsec:finite_mu_and_T}

\begin{figure}[htbp]
    \centering
    \includegraphics[width=7.8cm]{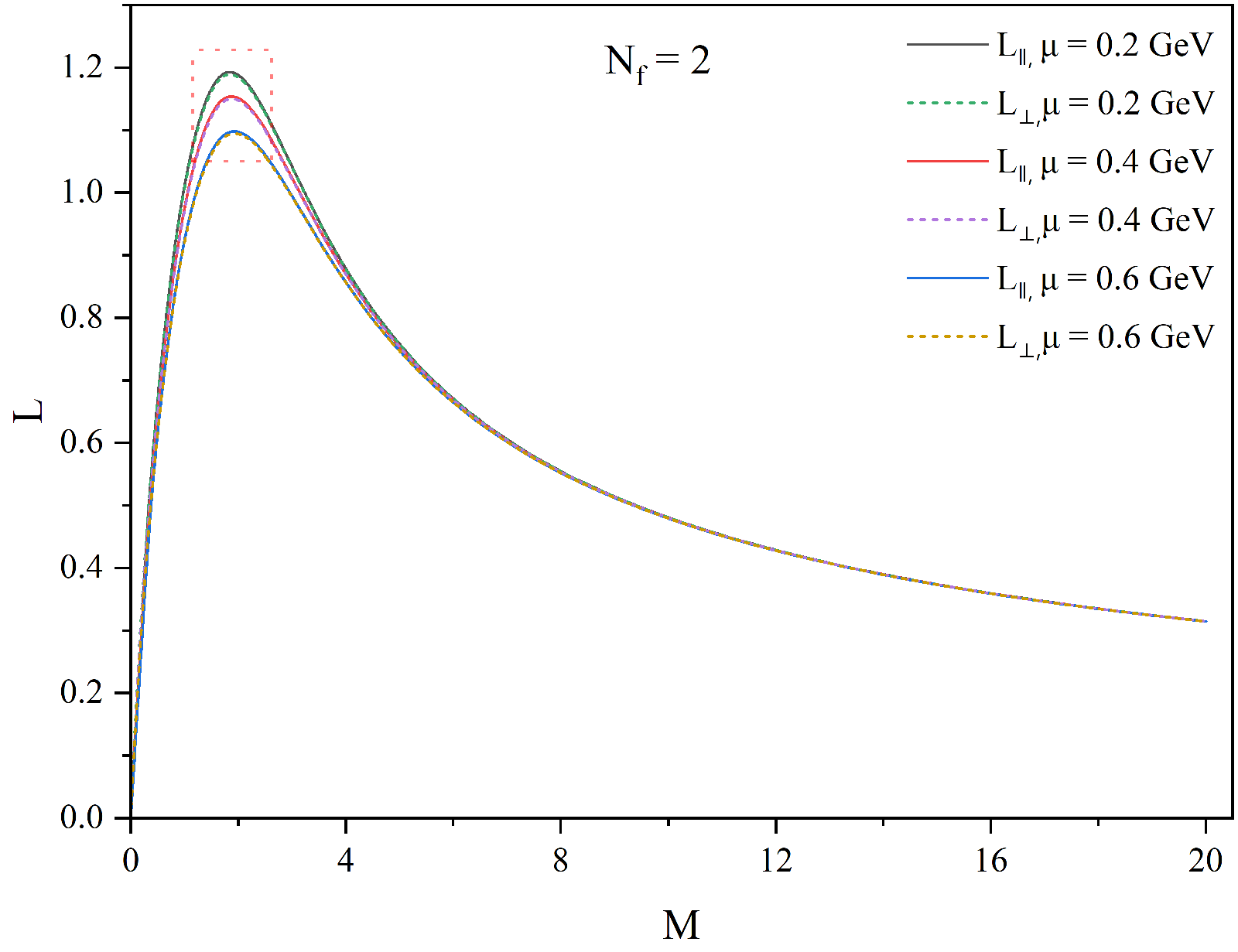}
    \includegraphics[width=7.8cm]{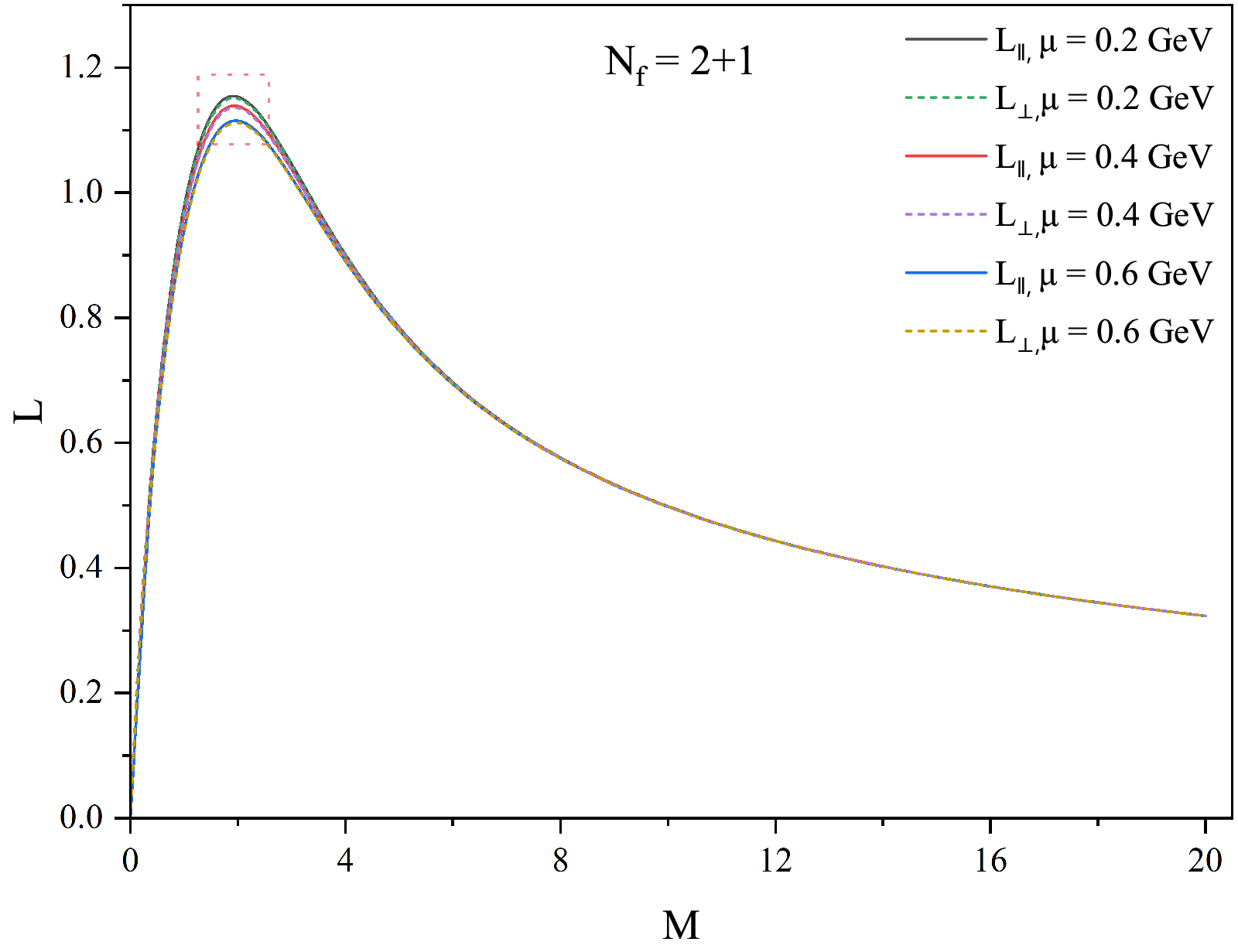}
    \caption{\(L\) versus \(M\) at \(\eta=0.1\) and
    \(T=0.3\ \text{GeV}\) for \(\mu=0.2,0.4,0.6\ \text{GeV}\).
    Solid: parallel; dashed: perpendicular. Left: \(N_f=2\); right:
    \(N_f=2+1\). The two-flavor peak at \(\mu=0.2\ \text{GeV}\) lies
    above the \(2+1\)-flavor peak, while the ordering reverses by
    \(\mu=0.6\ \text{GeV}\).}
    \label{fig2}
\end{figure}

\begin{figure}[htbp]
    \centering
    \includegraphics[width=7.8cm]{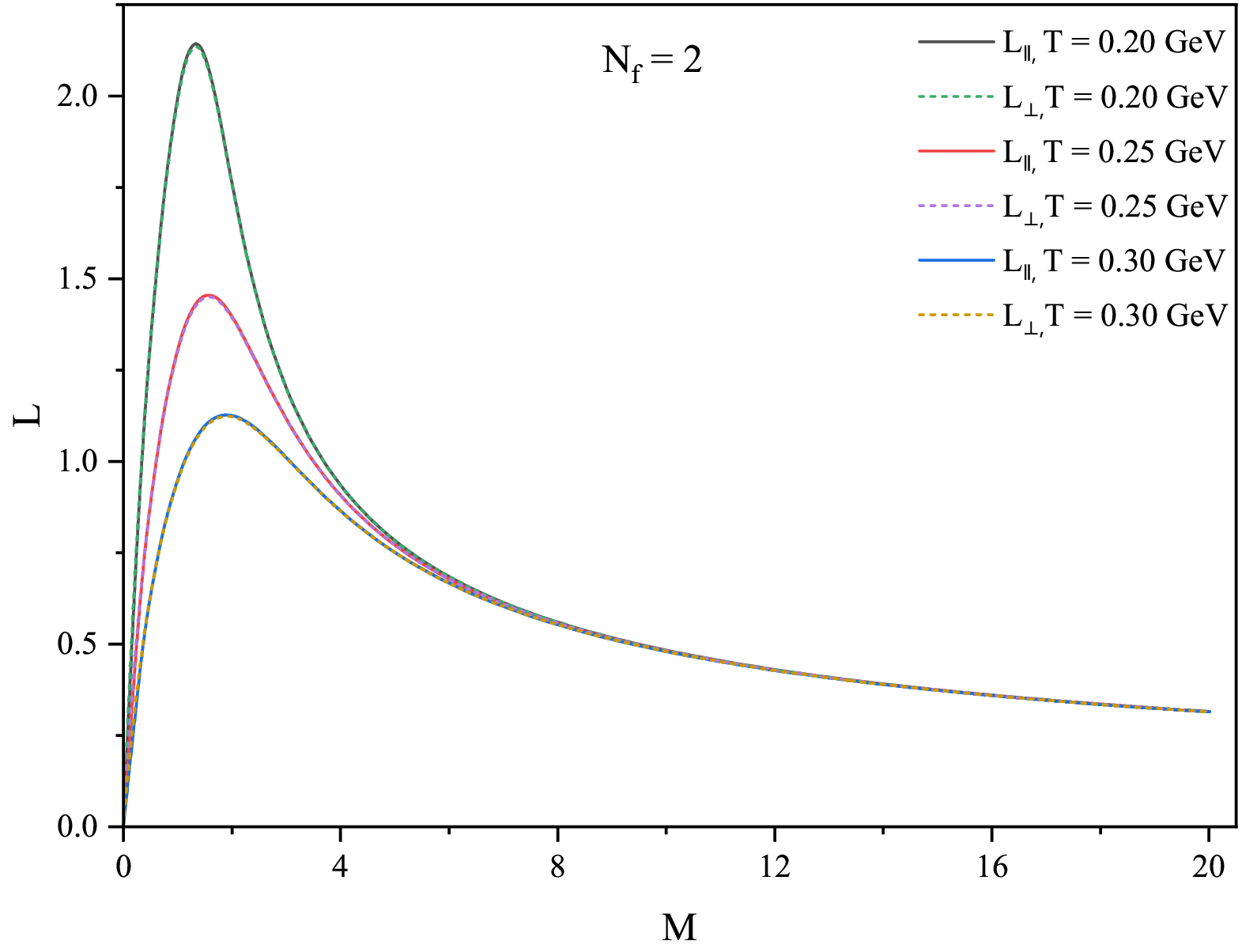}
    \includegraphics[width=7.8cm]{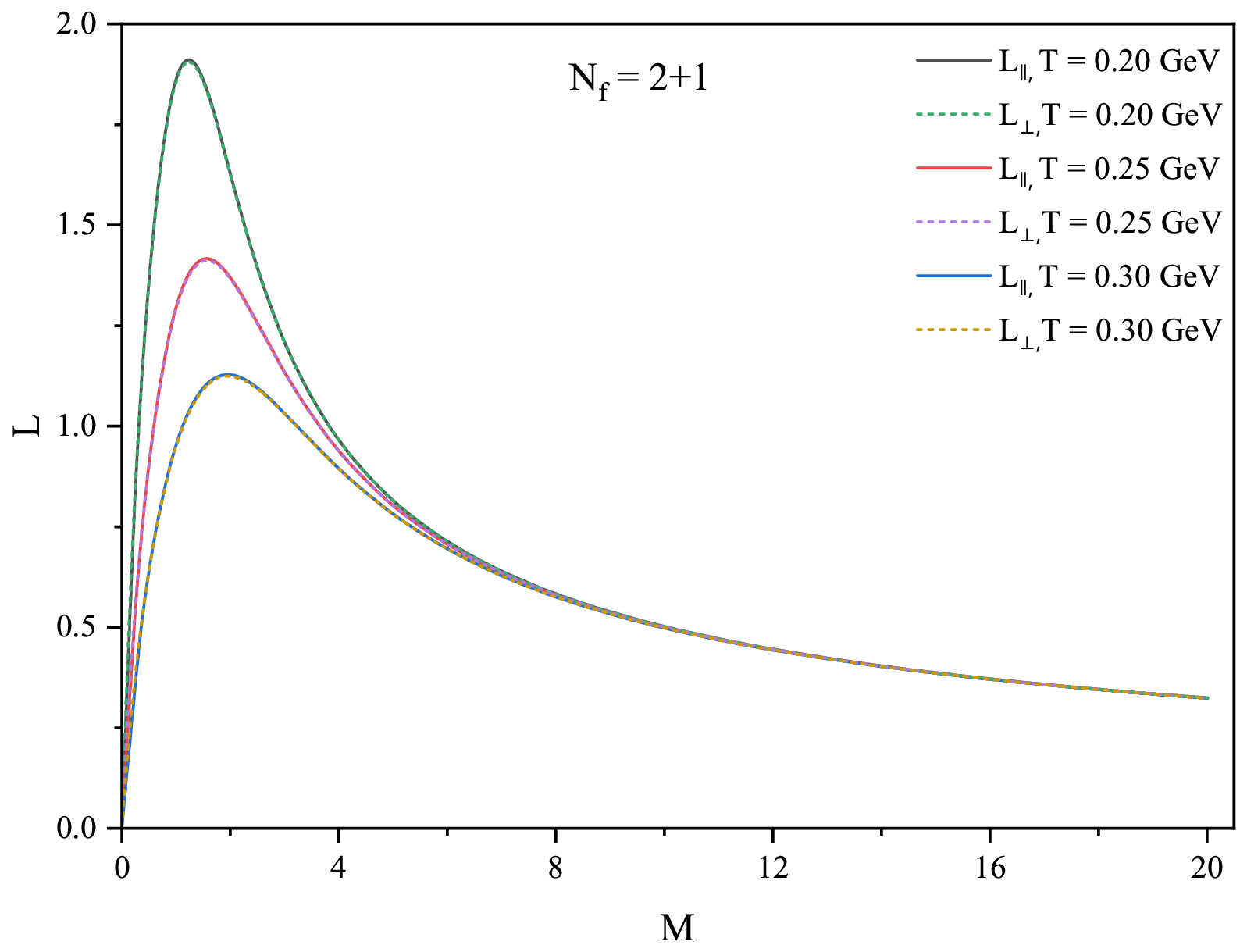}
    \caption{\(L\) versus \(M\) at \(\eta=0.1\) and
    \(\mu=0.5\ \text{GeV}\) for \(T=0.20,0.25,0.30\ \text{GeV}\).
    Solid: parallel; dashed: perpendicular. Left: \(N_f=2\); right:
    \(N_f=2+1\). Both flavor systems show the peak decreasing with
    \(T\); the splitting between parallel and perpendicular
    orientations remains at the percent level.}
    \label{fig3}
\end{figure}

Fig.~\ref{fig2} shows \(L(M)\) for \(N_f=2\) and \(N_f=2+1\) at
\(T=0.3\ \text{GeV}\) and \(\mu=0.2,0.4,0.6\ \text{GeV}\), and
Fig.~\ref{fig3} the same at \(\mu=0.5\ \text{GeV}\) and
\(T=0.20,0.25,0.30\ \text{GeV}\), both at \(\eta=0.1\). \(N_f=0\)
is omitted because \(c=0\) removes the density response.

Increasing \(T\) or \(\mu\) lowers the peak of \(L\) and reduces
\(L_s\). A higher temperature corresponds to a smaller horizon
radius \(z_t\), which restricts the maximal bulk depth and weakens
the binding. A larger \(\mu\) increases the magnitude of the
density contribution to \(g(z)\) through the \(c\mu^2 e^k\) term,
destabilizing the connected string. These trends are consistent
with the conformal \(L_s\propto1/T\) result
\cite{Liu:2007,Caceres:2006} and with previous holographic QCD
studies \cite{Zhu:2024,Zhang:2024}. The temperature trend is
qualitatively similar to Debye screening: a hotter medium screens
the inter-quark potential over a shorter distance. The density
trend reflects the additional baryon density in the calibrated
background.

The flavor ordering exhibits an inversion, visible in
Fig.~\ref{fig2}: at low \(\mu\) the \(N_f=2\) peak exceeds the
\(N_f=2+1\) peak, while at \(\mu\sim0.6\ \text{GeV}\) the two
cross. Two calibrated features produce this crossing. First,
\(T_c^{(N_f=2)}=0.189\ \text{GeV}\) is higher than
\(T_c^{(N_f=2+1)}=0.128\ \text{GeV}\), so at fixed absolute
temperature the two-flavor medium sits closer to its crossover and
its screening length is enhanced by effects near the crossover.
Second, \(|c_{N_f=2}|=0.377\) exceeds \(|c_{N_f=2+1}|=0.264\), so
the two-flavor screening length falls more steeply with \(\mu\).
The crossing is the point at which these two effects balance. Both
inputs are properties of the calibrated background rather than of
the Wilson loop itself. The crossing is therefore not a universal
statement about QCD, but a statement about the calibrated EMD
class that can be tested by an independent determination of the
flavor dependence of the static screening length.

\begin{figure}[htbp]
    \centering
    \includegraphics[width=7.8cm]{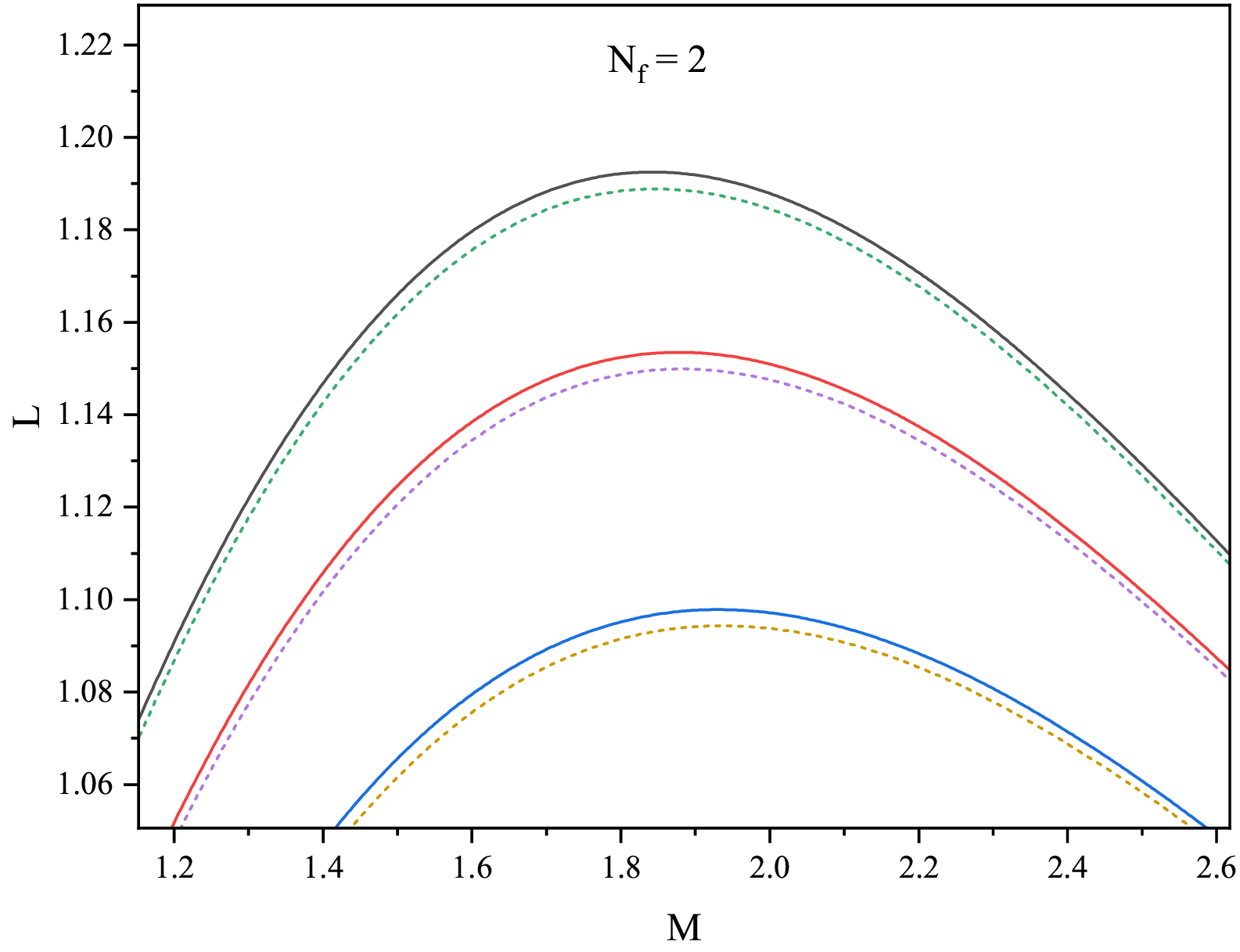}
    \includegraphics[width=7.8cm]{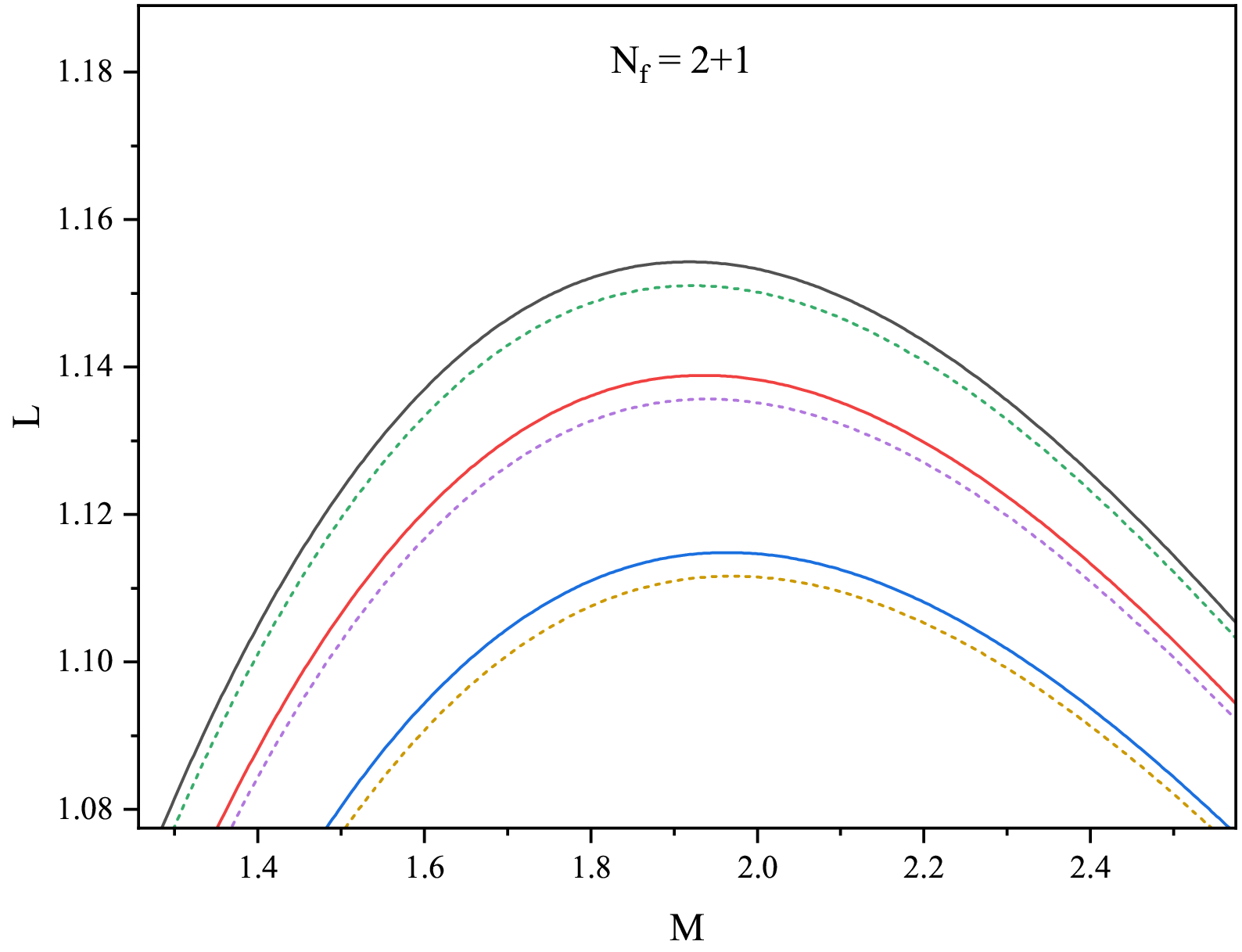}
    \caption{Enlarged peak region of Fig.~\ref{fig2}. Solid and
    dashed lines denote parallel and perpendicular orientations;
    parameters as in Fig.~\ref{fig2}. The solid curve peaks at a
    slightly larger \(L\) than the dashed curve at the same
    \((\mu,T,\eta)\).}
    \label{fig4}
\end{figure}

The parallel and perpendicular curves are nearly coincident at
\(\eta=0.1\) in Figs.~\ref{fig2} and~\ref{fig3}; the magnified
view in Fig.~\ref{fig4} resolves a percent-level splitting. This
is consistent with the weak anisotropy expected at low rapidity.
At \(\eta=0.3\) (Fig.~\ref{fig1}) the splitting is visibly
enhanced.

\begin{figure}[htbp]
    \centering
    \includegraphics[width=0.5\linewidth]{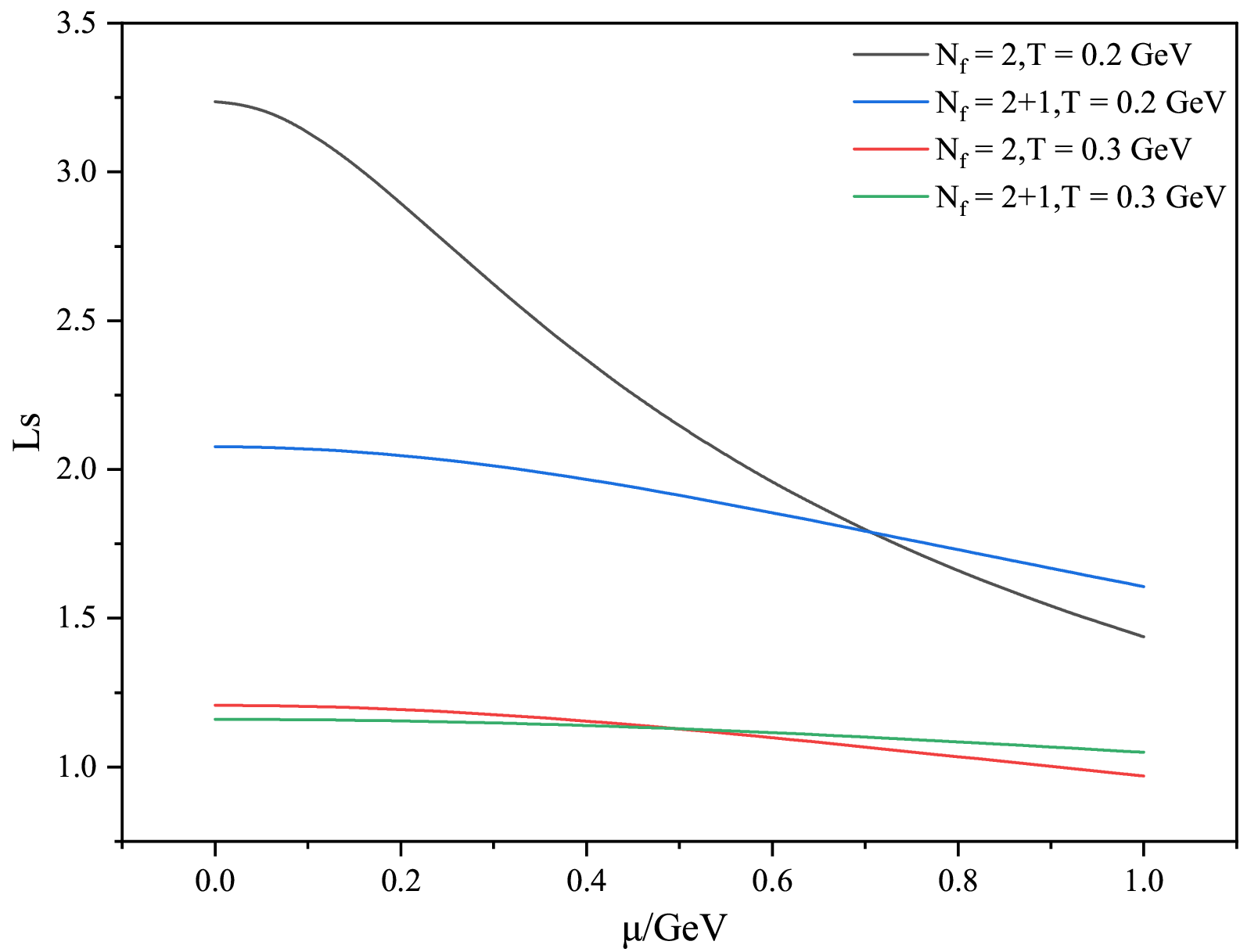}
    \caption{\(L_s\) versus \(\mu\) for different \(T\) and flavor
    systems at \(\eta=0.1\), parallel orientation. The two-flavor
    curve at \(T=0.2\ \text{GeV}\) decreases most steeply with
    \(\mu\), reflecting the larger \(|c|\) and the proximity to the
    two-flavor crossover.}
    \label{fig5}
\end{figure}

\begin{figure}[htbp]
    \centering
    \includegraphics[width=0.5\linewidth]{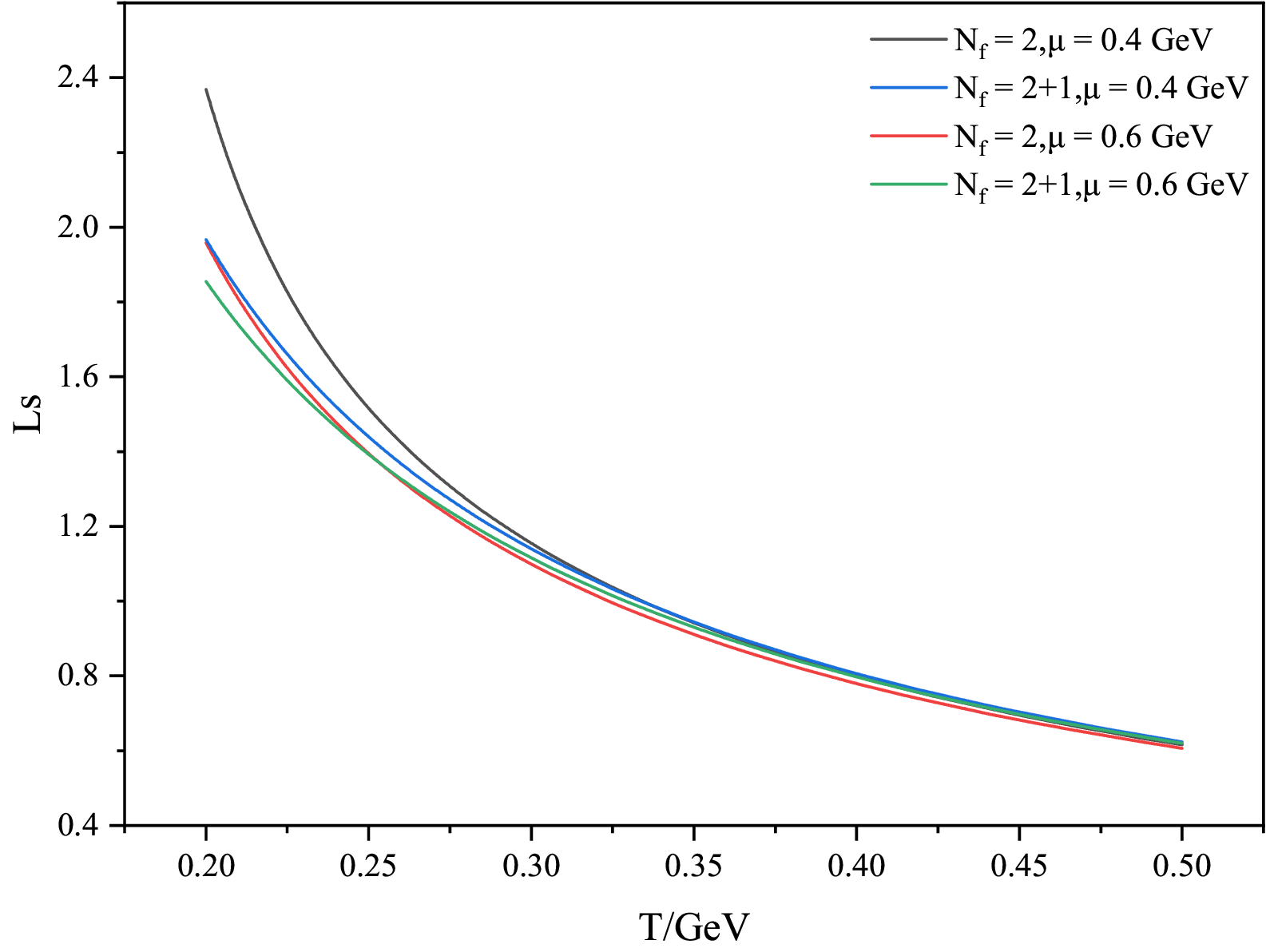}
    \caption{\(L_s\) versus \(T\) for different \(\mu\) and flavor
    systems at \(\eta=0.1\), parallel orientation. The curves
    approach one another at high \(T\) as the thermal term dominates
    the density contribution to \(g(z)\).}
    \label{fig6}
\end{figure}

The continuous dependence of \(L_s\) on \(T\) and \(\mu\) is shown
in Figs.~\ref{fig5} and~\ref{fig6}. In Fig.~\ref{fig6}, for
\(T\gtrsim0.4\ \text{GeV}\) the different flavor and \(\mu\)
curves are compressed within the displayed range. As \(T\) rises
and \(z_t\) decreases, the relative contribution of the \(\mu^2\)
term to \(g(z)\) is reduced by the thermal term, so the
sensitivity to flavor and density decreases. The curves are not
shown to converge to a universal scaling function; the differences
become small on the scale of the figure.

The sensitivity is largest at low \(T\). In Fig.~\ref{fig5}, the
\(N_f=2\) curve at \(T=0.2\ \text{GeV}\) falls much more steeply
with \(\mu\) than at \(T=0.3\ \text{GeV}\), indicating that the
density-driven modification of the blackening function is most
effective in the strongly nonconformal regime near the crossover.

\begin{figure}
    \centering
    \includegraphics[width=0.5\linewidth]{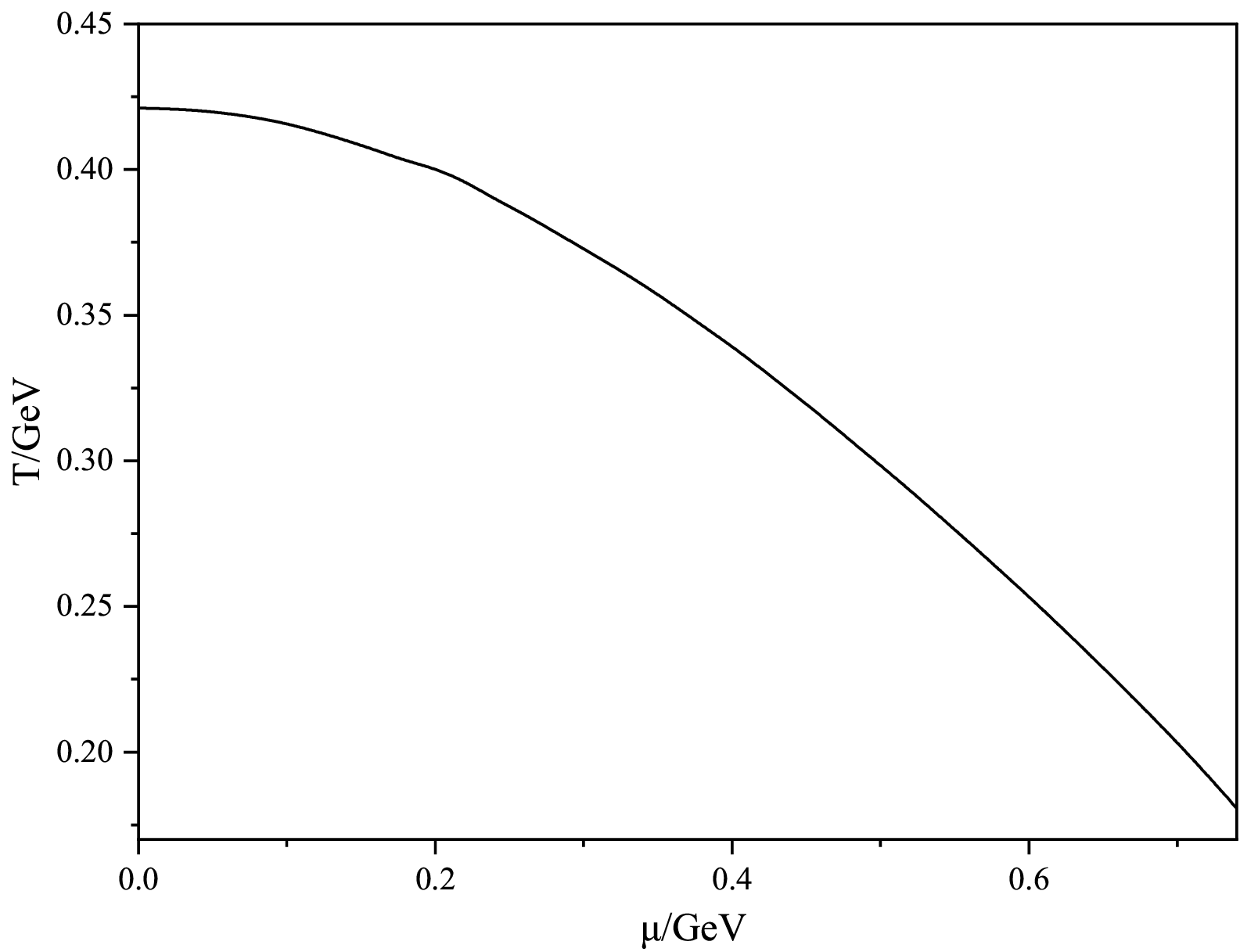}
    \caption{Flavor-inversion line in the \(T\)-\(\mu\) plane,
    defined by \(L_s^{(N_f=2)}=L_s^{(N_f=2+1)}\) at \(\eta=0.1\).
    Below the curve \(L_s^{(N_f=2)}>L_s^{(N_f=2+1)}\); above it the
    ordering is reversed. The line terminates when \(T\) falls below
    the two-flavor crossover.}
    \label{fig7}
\end{figure}

We define the flavor-inversion line by
\(L_s^{(N_f=2)}=L_s^{(N_f=2+1)}\). In Fig.~\ref{fig7} it falls
monotonically from \(T=0.421\ \text{GeV}\) at \(\mu=0\) to
\(T\approx0.18\ \text{GeV}\) at \(\mu\approx0.72\ \text{GeV}\).
The endpoint occurs where the temperature drops below
\(T_c^{(N_f=2)}\), beyond which the deconfined description is not
applicable. On the low-temperature side of the line the
enhancement near the crossover of the two-flavor system dominates;
on the high-temperature side the larger density sensitivity of the
two-flavor medium reverses the ordering. The crossing requires the
two-flavor crossover temperature to exceed the \(2+1\)-flavor one
while the two-flavor density-coupling magnitude also exceeds the
\(2+1\)-flavor value; in the present parameter set this condition
is satisfied. The precise location is tied to the fitted
parameters and should be refined by propagating the posterior. The
negative slope reflects the fact that the density-driven mechanism
grows with \(\mu\) while the enhancement near the crossover is
strongest at low \(T\).

\begin{figure}[htbp]
    \centering
    \includegraphics[width=7.8cm]{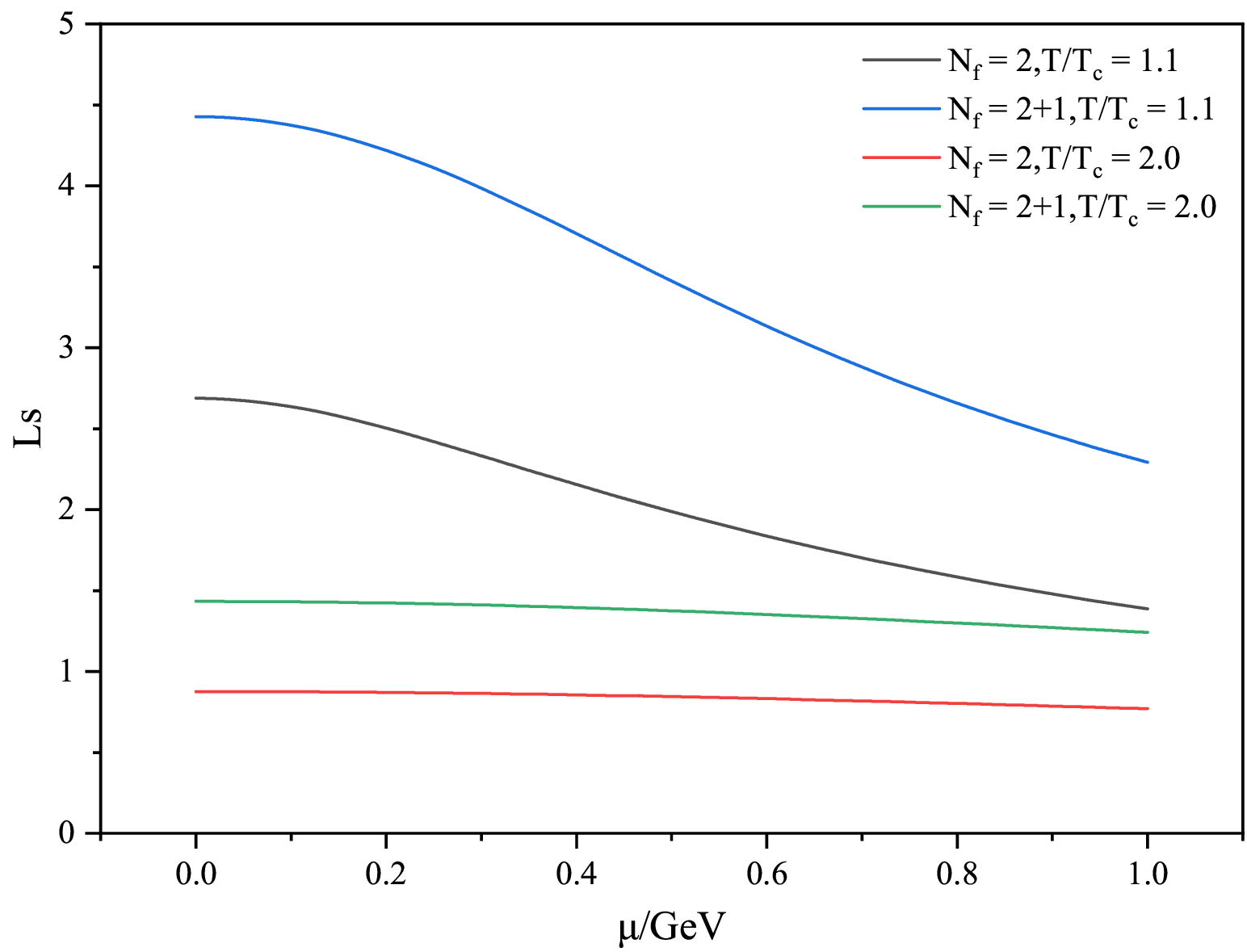}
    \includegraphics[width=7.8cm]{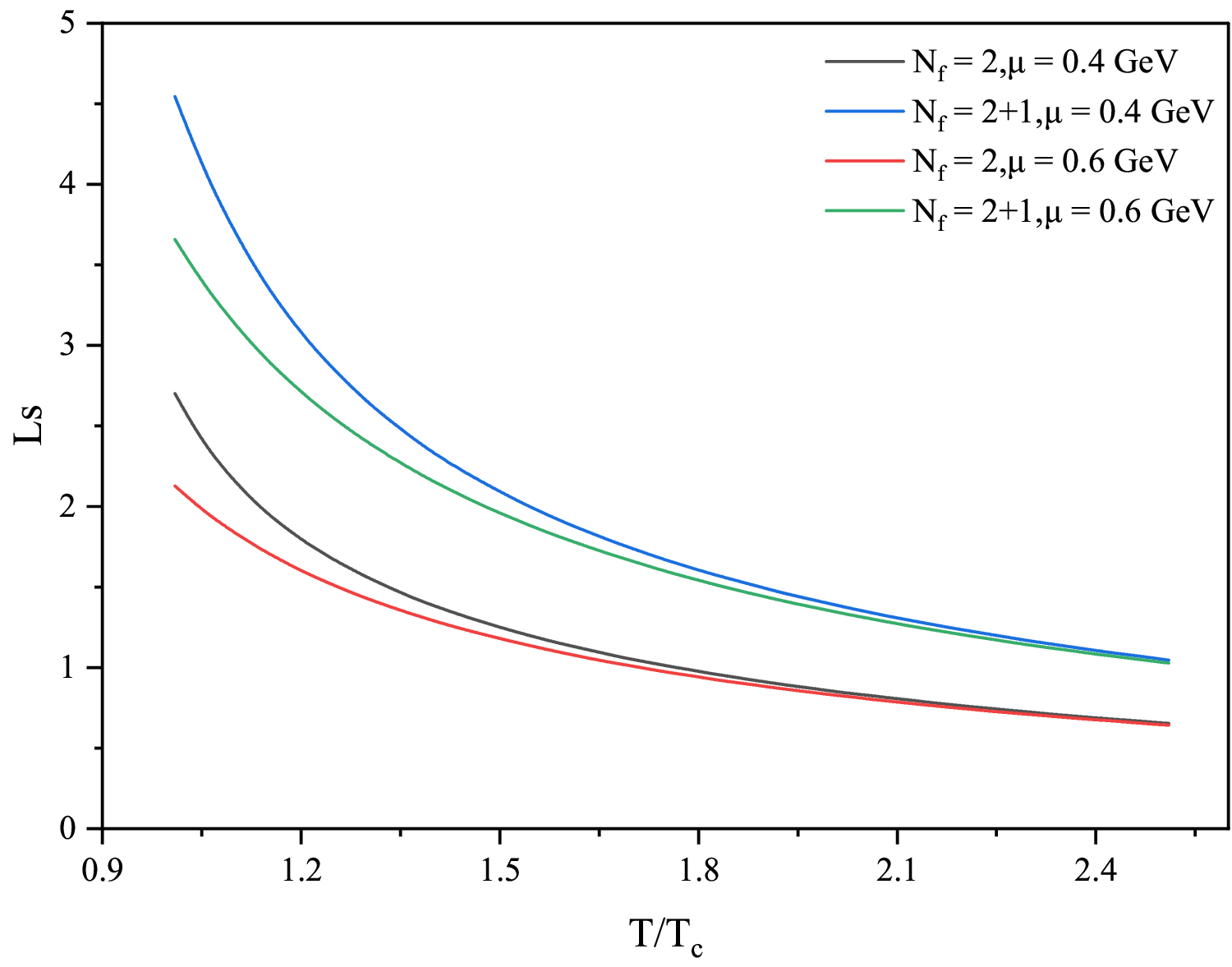}
    \caption{\(L_s\) in the reduced-temperature plane at
    \(\eta=0.1\), parallel orientation. Left: \(L_s\) versus \(\mu\)
    at fixed \(T/T_c=1.1\) and \(2.0\). Right: \(L_s\) versus
    \(T/T_c\) at fixed \(\mu=0.4\) and \(0.6\ \text{GeV}\). The
    \(2+1\)-flavor curve lies above the two-flavor one throughout;
    the inversion of Fig.~\ref{fig7} is absent in this mapping
    because a fixed \(T/T_c\) corresponds to different absolute
    temperatures for the two flavor systems.}
    \label{fig8}
\end{figure}

Fig.~\ref{fig8} shows \(L_s\) in the reduced-temperature plane
\(T/T_c\). Here the \(2+1\)-flavor curve lies above the two-flavor
one throughout, with no inversion. This is an absolute-temperature
mapping effect, not a universal flavor ordering: because
\(T_c^{(N_f=2+1)}<T_c^{(N_f=2)}\), a fixed \(T/T_c\) corresponds
to a lower absolute temperature for the \(2+1\)-flavor system, and
since \(L_s\) falls steeply with absolute \(T\), the
\(2+1\)-flavor curve is displaced upward. The absence of inversion
in this plane shows that the inversion of Fig.~\ref{fig7} is a
statement about comparing flavor systems at the same absolute
temperature while their crossover temperatures differ.

\subsection{Model dependence and robustness}
\label{subsec:robustness}

The results fall into two classes. For the flavored systems, the
monotonic decrease of \(L_s\) with \(T\), \(\mu\), and \(\eta\),
together with \(L_{s\perp}<L_{s\parallel}\), follows from the
horizon structure and the wind-induced effective geometry. These
features require only that the horizon move toward the boundary as
\(T\) or \(\mu\) increases and that \(h(z)<g(z)\) for \(\eta>0\);
they do not depend on the detailed fitted parameters. The
zero-density flavor ordering and the inversion line depend on the
calibrated crossover temperatures, the fitted warp factors, and
the density coupling \(c\). The inversion line is therefore a
prediction of this EMD calibration rather than a model-independent
QCD statement. Its existence requires a specific sign combination
of the \(T_c\) and \(c\) differences; its location should be
replaced by a credible band once the parameter posterior is
propagated.

This distinction determines how each result should be tested. A
qualitative statement such as \(L_{s\perp}<L_{s\parallel}\) can be
examined in any holographic background of this type with a horizon
and a nontrivial wind. A quantitative statement such as the value
of \(\mu\) at which the two-flavor and \(2+1\)-flavor curves cross
requires the specific calibration used here. The monotonic
dependences and the orientation ordering are therefore the stable
part of the result, and the inversion line is the
calibration-dependent part. The flavor ordering at zero density
has a direct lattice interface: the static \(L_s\) for \(N_f=2\)
versus \(N_f=2+1\) can be compared once the flavor dependence of
the static free energy is under control, following the strategy of
Ref.~\cite{Kaczmarek:2005}. The inversion line has an experimental
interface through beam-energy-scan trajectories in the
\(T\)-\(\mu\) plane \cite{Andronic:2016,Andronic:2024,Veres:2019};
the statement to be tested is the existence and negative slope of
the line, not its numerical position at the current fitted
calibration.

We also compare the orientation dependence with the noncommutative
and magnetic results of Refs.~\cite{Wu:2024,Wu:2025}. In
Ref.~\cite{Wu:2024}, noncommutativity was reported to increase the
screening length and reduce quarkonium dissociation, with the
effect most visible when the \(Q\bar{Q}\) axis lies along a
noncommutative direction. In Ref.~\cite{Wu:2025}, a magnetic field
reduces \(L_s\) and enhances dissociation; the smallest \(L_s\)
occurs when both the boost and the pair axis are perpendicular to
the magnetic field. In the present EMD background there is no
intrinsic spatial anisotropy, so the wind is the only anisotropic
source and the parallel/perpendicular ordering is controlled by
the horizon structure. The noncommutative and magnetic
calculations show that intrinsic anisotropy can change \(L_s\) and
its orientation dependence in ways that depend on the origin of
the anisotropy. A direct numerical comparison would require
matching the anisotropy source and the thermodynamic trajectory,
which we do not attempt here.

\section{Conclusion}
\label{sec:conclusion}

We have computed the screening length of a moving heavy
quark-antiquark pair in a lattice-calibrated EMD background for
pure-glue, two-flavor, and \(2+1\)-flavor QCD, at finite \(T\), at
finite \(\mu\) for the flavored systems, for parallel and
perpendicular orientations, and for \(\eta=0.1\) and \(0.3\). This
calculation complements the jet-quenching calculations on the same
background \cite{Zhang:2025,Zhu:2026}. Unlike
Refs.~\cite{Wu:2024,Wu:2025}, the present background has no
intrinsic spatial anisotropy; the wind is the only source.
References~\cite{Wu:2024,Wu:2025} show that intrinsic anisotropy
can change \(L_s\) and its orientation dependence, whereas the
present calculation isolates the wind and flavor calibration. A
direct comparison would require matching the anisotropy source and
the thermodynamic trajectory.

The main results are as follows. \(L_s\) decreases monotonically
with \(T\) and \(\eta\) in every flavor configuration and
orientation; for the flavored systems it also decreases
monotonically with \(\mu\). The perpendicular screening length is
systematically smaller than the parallel one. At zero density and
\(T=0.3\ \text{GeV}\), in this calibration additional dynamical
flavors shorten \(L_s\). The monotonic and orientation statements
follow from the horizon structure and the wind-induced effective
geometry; the zero-density flavor ordering reflects the calibrated
background. The parallel and perpendicular orientations are the
two principal axes of the wind-induced anisotropy; because the EMD
medium is transversely isotropic, all perpendicular directions are
equivalent and the two limits capture the independent orientation
dependence. A full angular scan would refine the interpolation but
is not needed for the ordering and monotonicity statements.

The calibration-dependent elements are the numerical position of
the flavor-inversion line and the percent-level size of the
parallel and perpendicular splitting at low rapidity. The
inversion line separates a regime in which the two-flavor \(L_s\)
exceeds the \(2+1\)-flavor one from a regime in which the ordering
is reversed, and it terminates when the temperature falls below
the two-flavor crossover. Its appearance requires
\(T_c^{(N_f=2)}>T_c^{(N_f=2+1)}\) together with
\(|c_{N_f=2}|>|c_{N_f=2+1}|\); in the present parameter set both
conditions hold. Its location is a fitted prediction of this
calibration and should be replaced by a credible band once the
parameter posterior is propagated.

Several extensions are possible. Propagating the EMD parameter
posterior would turn Fig.~\ref{fig7} into a band and quantify the
model uncertainty. Extending the calculation to arbitrary
orientation angles and to higher rapidities would allow a
determination of the effective exponent in
\(L_s\propto(1-v^2)^{\Delta}\), to be compared with the conformal
\(\Delta=1/4\). A deviation from \(1/4\) would probe nonconformal
effects. On the experimental side, the inversion line can be
mapped onto beam-energy-scan trajectories and compared with
charmonium and bottomonium suppression patterns
\cite{Andronic:2016,Andronic:2024,Veres:2019}. On the lattice
side, the static flavor dependence of \(L_s\) provides a direct
check of the ordering at zero density. Incorporating finite quark
mass and string backreaction would bring the Wilson-loop
dissociation criterion closer to that of physical quarkonia, for
which the \(Q\bar{Q}\) separation is comparable to the string
scale.

\section*{Acknowledgments}
This work is supported by the National Natural Science Foundation
of China under Grant No.~12375140 and the National Key Research
and Development Program of China with Grant No.~2024YFA1610804,
the Fundamental Research Funds for National Universities, China
University of Geosciences No.~2025XLB102.

\end{document}